\documentclass[review]{elsarticle}

\usepackage{lineno,hyperref}
\usepackage{amsmath}
\modulolinenumbers[5]

\begin{document}

\begin{frontmatter}

\title{Laboratory light scattering from regolith surface and simulation of data by Hapke model}

\author[SD]{S. Deb}
\ead{debsanjeeb@gmail.com}
\author[SD]{A. K. Sen}
%\ead{asokesen@yahoo.com}
\address[SD]{Department of Physics, Assam University, Silchar, Assam, India, Pin-788011}
%%\fntext[myfootnote]{Since 1880.}




\begin{abstract}
The small atmosphereless objects of our solar system, such as asteroids, the moon are covered by layer of dust particles known as regolith, formed by meteoritic impact. The light scattering studies of such dust layer by laboratory experiment and numerical simulation are two important tools to investigate their physical properties. In the present work, the light scattered from a layer of dust particles, containing 0.3$\mu$m $Al_{2}O_{3}$ at wavelength 632.8 nm is analysed. This work has been performed by using a light scattering instrument 'ellipsometer', at the Department of Physics, Assam Universiy, Silchar, India. Through this experiment, we generated in laboratory the photometric and polarimetric phase curves of light scattered from such a layer. In order to numerically simulate this data, we used Hapke's model combined with Mie's single particle scattering properties. The perpendicular and parallel components of single particle albedo and the  phase function were derived from Mie theory. By using the Hapke's model combined with Mie theory,  the physical properties of the dust grain such as grain size, optical constant $(n,k)$ and wavelength can be studied through this scheme. In literature, till today no theoretical model to represent polarisation caused due to scattering from rough surface is available, which can successfully explain the scattering process. So the main objective of this work is to develop a model which can theoretically estimate polarisation as caused due to scattering from rough surface and also to validate our model with the laboratory data generated in the present work.
\end{abstract}


\end{frontmatter}

%\linenumbers

\section{Introduction}
The unpolarised solar radiation scattered by asteroidal regolith surface (a particulate medium) is partially polarised (Hapke 1993).  The polarisation phase curve, i.e. the variation of polarisation (P) as a function of phase angle $(\alpha)$ provides the main source of information about the physical properties of regolith, such as grain size, optical constant, albedo, surface roughness and porosity. The polarimetric parameters such as maximum and minimum polarisation; the phase angles corresponding to these extrema; inversion angle and slope at inversion can be retrieved from the polarisation phase curve. In order to account for different branches of polarisation phase curve, semiemperical relationships have been developed between polarisation values (such as maximum polarisation $P_{max}$ , slope at inversion) and albedo of the scattering medium (Umov 1905, KenKnight et al. 1967, Widorn 1967, Zellner 1977).  Due to observational constrain, asteroids can only be observed in limited phase angle (Shestopalov 2004). In laboratory simulation, this constrain can be eliminated and observation can be carried out at different phase angles. The laboratory simulation obtained for layer of dust particles with multiple scattering, can be compared with remote sensing data of the moon and asteroids to retrieve the physical properties of regolith present on them.

The present simulation work is being reported from the same laboratory, where in past the Hapke bidirectional reflectance model (Hapke 1993) has been widely used for modeling the photometric laboratory data (Deb et al. 2011, Bhattacharjee et al. 2011, Deb et al. 2012). A specific single particle scattering phase function p(g) is needed in Hapke model. The regolith consists of irregular particles, whose single particle phase function is generally unknown. So empirical phase functions such as one term or two term Henyey Greenstein phase function is incorporated in Hapke model. But the empirical phase functions do not explicitly contain the physical parameter, such as grain size and optical constant. In order to directly study the effects of these parameters, an exact theoretical single scattering phase function of these parameters is required. The exact solution of scattering by a smooth and homogeneous sphere is provided in Mie theory (Mie 1908).  The solution of Mie theory depends on single particle properties such as grain size, optical constant and also on wavelength of incident radiation. So by incorporating a phase function $p(g)$ from Mie theory in Hapke Model, the effect of physical parameter can be studied.

The Hapke model provides good agreement with laboratory data. Deb et al. 2011 used Hapke model (Hapke 1993) in combination with Mie theory to study the bidirectional reflectance of 0.3 $\mu$m alumina at 632.8 nm. The author found the best fit between the experimental data and model curve for absorption coefficient k=0.000009 for Alumina particles, which was in close agreement with the value obtained in Piatek et al. 2004. Deb et al. 2011 also compared the variation of relative intensity with particle size of alumina samples at 13 different diameters ranging from 0.1 to 30.09 $\mu$m (Nelson et al. 2000), with the model curve for absorption coefficient k=0.000009, based on above method. The close fit in above comparison confirmed the correctness of method used by Deb et al. 2011.

Deb et al. 2012 used the same method as described in Deb et al. 2011, for fitting the variation of relative intensity with particle size of a given sample ( containing basalt and olivine) (Kaasalainen et al. 2003) and dunite particles (Kamei et al. 2002). For this given sample, Deb et al. 2012 also obtained a satisfactory fit between an empirical relation $I=D^{-a}+b$ ($D$=diameter of dust grain) and relative intensity variation with particle size, with suitable values of the parameters a and b. The authors concluded that the average particle size of actual regolith with known composition can be estimated by using this empirical relation.

Bhattacharjee et al. 2011 studied the bidirectional reflectance of 0.3 $\mu$m and 1.0  $\mu$m alumina at two wavelengths 543.5 nm and 632.8 nm. The author used slightly different approach from Deb et al. 2011, 2012. The single particle albedo $w$ and asymmetry parameter $\xi$ were computed from Mie theoy. The asymmetry parameter was used to calculate an empirical phase function, Known as Henyey-Greenstein phase function $p(g)$. The authors theoretically computed the bidirectional reflectance of alumina samples by using $w$ and $p(g)$ in Hapke model (Hapke 1993). For fitting the bidirectional reflectance data with model curve, the absorption coefficient k was used as a free fitting parameter. The authors obtained best fitted values of k as 0.000001 and 0.00001 at 543.5 nm and 632.8 nm respectively. The value k=0.000001 at 543.5 nm was comparable with the value k=7.86$\times10^{-6}$ at 635 nm obtained in Piatek et al. 2004.

In addition to the above photometric work, some researchers have also worked on the polarisation resulting from rough surface scattering.

 Hadamcik et al. 1996,2002,2003 had investigated the polarisation phase curves of regolith analogue of different terrestrial, meteorite and synthetic  samples by using $PROGRA^{2}$ experimental setup. The variation of degree of polarisation with phase angle for basalt of different grain size range was obtained by Hadamcik et al. 1996. It was found that the value of maximum polarisation $P_{max}$ increases with grain size and corresponding phase angle $\alpha_{max}$ decreases with grain size. For both fluffy and compact particles, Hadamcik et al. 2002 found that the variation of $P_{max}$ as fuction of albedo follows the Umov law in case of both sifted (just sieved and not pressd) and packed (the surface is pressed with spatula) samples. The authors also obtained the wavelength dependence of $P_{max}$ for samples with two sets of fluffy particles and concluded that this fluffy particles could be good candidate for cometary dust. Hadamcik et al. 2003 found that the polarisation decreases with increase in interaction between the glass beads from single bead to agglomerates of beads and finally  to deposited beads, which is due to the increase in multiple scattering.

Worms et al. 1996,1999a,1999b had done the polarimetric laboratory simulation of deposited sample by $PROGRA^{2}$ experimental setup. The polarisation phase curve of 8-10 $\mu$m and  11-15 $\mu$m sized deposited boron carbide particles in sifted and packed  form, were studied by  Worms et al. 1996. It was found that in both size ranges, the polarisation is higher in packed sample than in sifted sample and also polarisation is higher for bigger grains as compared to smaller grains. For deposited samples of basaltic glass with grain sizes 10$\mu$m and 45$\mu$m in sifted and packed form, Worms et al. 1999a found that $P_{max}$  is higher for packed samples than sifted samples. Worms et al. 1999b noticed for $SiC$ and $B_{4}C$ samples, $P_{max}$ increases with increase in grain size, which confirmed the findings of Worms et al. 1996 and Hadamcik et al. 1996. It was also found that for 9 $\mu$m $B_{4}C$, 7 $\mu$m and  13 $\mu$m $SiC$ deposited samples, $P_{max}$ is higher for packed as compared to sifted sample.

In Penttila et al. 2003, the authors theoretically analysed the polarimetric data of (9$\pm$1) $\mu m$,(13$\pm$2) $\mu m$ and (88$\pm$5) $\mu m$ boron carbide ($B_{4}C$) generated by  $PROGRA^{2}$ microgravity experiment. For theoretical analysise, authors used a new model to describes the shapes of stochastic polyhedra with ray-tracing code and obtained (n,k)=(2,0.04), as the best fitted optical constant for boron carbide.

Mcguire et al. 1995 experimentally studied the light scattering by large  spherical and irregular particles (approximately 1 cm in size). The authors found that for phase angle between $30^{0}$ and $70^{0}$, the difference (between two orthogonal components of polarisation) $\Delta I(g)$ of scattered radiation from the particles is a proportional to the difference between Fresnels reflection coefficients. These coefficients depend only on refractive index for dielectric material. The authors concluded that this method may be applied to determine average value of refractive index of regolith present on solar system objects.

In literature, a large amount of laboratory gererated polarimetric data on scattering from rough surface are available from experiment (Hadamcik et al. 1996,2002,2003; Worms et al. 1996,1999a,1999b). For simulating these data, no appropriate theoretical model which include the physical properties of the constituent particles, is available. In the present work, a model has been developed based on Mie theory (Mie 1908) and Hapke model (Hapke 1993) to describe the polarisation of the light scattered from rough surface, in terms of physical properties of the constituent particles. For verifying the model's correctness, the model curve is compared with laboratory data generated in the present work by the instrument ellipsometer.

\begin{figure*}[!htbp]
\centering
\includegraphics{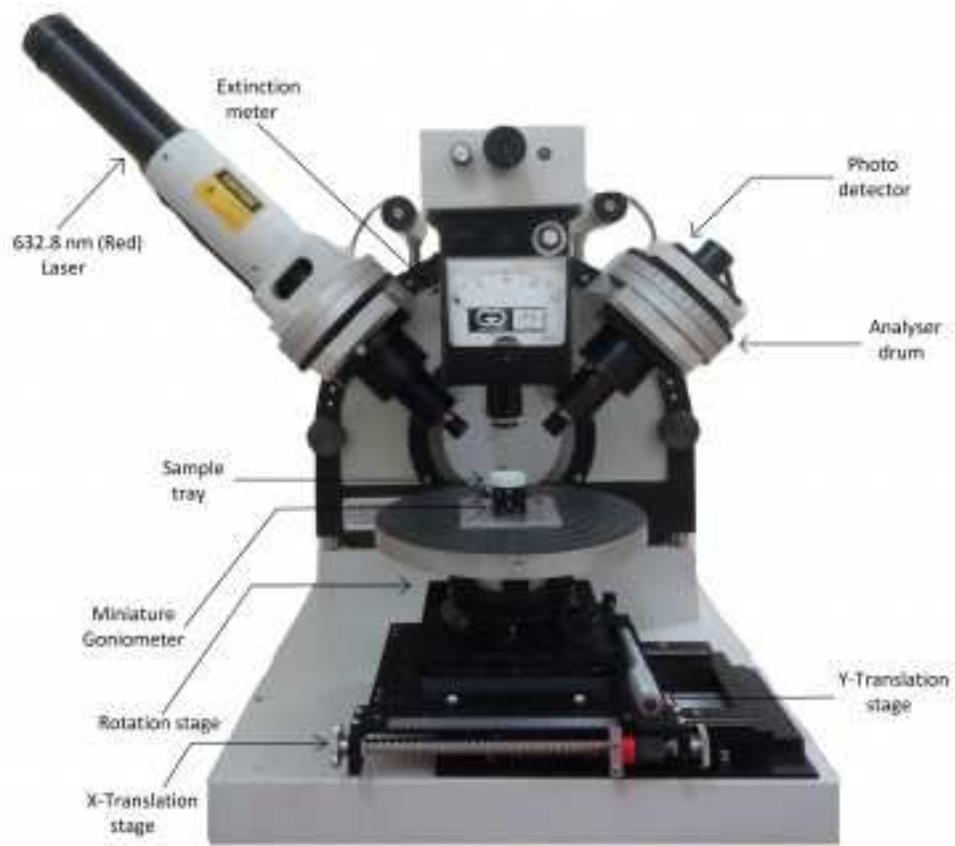}\\
\caption{Photograph of L117 Gaertner Ellipsometer showing different components.}
\end{figure*}

\section{Instrumental details and sample preparation}
The instrument 'L117 Gaertner ellipsometer' is a polarimeter used in the present work to simulate the process of light scattering from regolith surface. The phase angle dependence of reflectance and polarisation of scattered ligth can be estimated by this instrument. The ellipsometer with different components is shown in fig.1. The light source in the ellipsometer is a linearly polarised He-Ne laser having wavelength 632.8 nm (red) and beam diameter 15$\mu m$. A photo diode is used as a detector to measure the intensity of scattered light from sample surface. A rotatable glan-thompson prism attached in front of the detector is used as an analyser. In the ellipsometer, samples are placed horizontally on a sample tray. The light source and the detector are rotated in a vertical scattering plane to produce different scattering geometries. The angle of incidence and emergence in ellipsometer can be varied from $35^{o}$ to $70^{o}$ in the steps of $5^{o}$, so that the phase angle coverage is $70^{o}$-$140^{o}$. In the present case, two different sets of scattering geometries are used to produce the phase curves, viz keeping incident angle fixed at i=$45^{o}$ and $60^{o}$, the emergent angle is varied from $35^{o}$ to $70^{o}$; and keeping emergent angle fixed at e=$35^{o}$, $45^{o}$, $50^{o}$, $55^{o}$ and $60^{o}$,the incident angle is varied from $35^{o}$ to $70^{o}$. The powdered material of 0.3$\mu$m sized $Al_{2}O_{3}$ is used as sample to produce regolith surface. The sample is poured gently in a sample tray. The top surface of sample is pressed with a glass spatula to make it smooth.

\section{Generation of photometric and polarimetric data}

For the polarimetric measurement, a procedure is followed from the work of Sen et al. 1900 and Clarke 1971. The authors used this method for cometary polarimetric measurements. In the present case, this method is used for laboratory polarimetric measurements. The intensities $I_{1}$, $I_{2}$ and $I_{3}$ of the scattered light from the sample surface at three orientation $0^{0}$,$120^{0}$ and $240^{0}$ of the glan-thompson are recorded at each phase angle. From these intensities, the actual scattered intensity and the polarisation can be calculated by the relations,

\begin{subequations}
\begin{align}
 I &=\frac{2}{3}(I_{1}+I_{2}+I_{3})\label{1a}\\[10pt]
 P_{observe} &=\frac{2 \sqrt{I_{1}(I_{1}-I_{2})+I_{2}(I_{2}-I_{3})+I_{3}(I_{3}-I_{1})}}{I_{1}+I_{2}+I_{3}}\label{1b}
\end{align}
\end{subequations}

Since the polarisation $P_{observe}$ values are expressed as ratios of intensities, so no absolute photometric callibration is required for our measurements.

\section{Modeling of normalised reflectance and polarisation}
The well known reflectance model by Hapke 1993 is mostly used to interpret the laboratory simulated bidirectional reflectance data. To theoretically estimate the reflectance, we shall follow here an approach as described in our previous work Deb et al. 2011, 2012. The model requires two unknown parameters viz. the opposition surge amplitude $B_{0}$ and opposition angular width $h$, which are important mainly at phase angle smaller than $15^{0}$. Their effect can be neglected at large phase angles. This model also requires another parameter, single particle albedo \textit{w} and a single particle phase function $p(g)$. The values of \textit{w} and $p(g)$ can be obtained by using Mie theory  as a function of three physical parameters viz. grain size of sample, optical constant $(n,k)$ of the sample and monochromatic wavelength of incident radiation. With the help of Mie theory, one can introduce \textit{w} and $p(g)$ in Hapke model, and thus the physical properties of the grains can be included for numerical modeling. The formula for bidirectional reflectance as described by Hapke 1993 is reproduced below:

\begin{equation}\label{3}
 r(i,e,g)=\frac{1}{4\pi}(\frac{\mu_{o}}{\mu+\mu_{o}})\Big\{[1+B(g)]\textit{w}p(g)+\textit{w}[H(\mu_{o})H(\mu)-1]\Big\}
\end{equation}
where $\mu_{0}=\cos i$ and $\mu=\cos e$,

$B(g)=\frac{B_{0}}{1+(\frac{1}{h})tan (\frac{g}{2})}$

$B_{o}$= opposition surge amplitude,  $h$= opposition surge width

 \textit{w} = single-particle scattering albedo, $p(g)$= single particle scattering phase function at phase angle $g$ and in the present study Mie phase function is used,
\begin{equation}\label{4}
H(x)=\Big[1-(1-\surd(1-\textit{w}))x\{r_{o}+(1-\frac{1}{2}r_{o}-r_{o}x)\}\ln\frac{1+x}{x}\Big]^{-1}
%H(x)=\frac{1+2x}{1+2\gamma x}
\end{equation}
and
\begin{equation}\label{5}
r_{o}=\frac{2}{1+\surd(1-\textit{w})}-1
\end{equation}
As described in the present work the phase angle is always greater than $15^{0}$, so the the opposition effect can be neglected (Hapke 1993). Using $B(g)$=0 in Equation (\ref{3}), we have
\begin{equation}\label{6}
 r(i,e,g)=\frac{1}{4\pi}(\frac{\mu_{o}}{\mu+\mu_{o}})\Big\{\textit{w}p(g)+\textit{w}[H(\mu_{o})H(\mu)-1]\Big\}
\end{equation}
From Equation (\ref{6}), the perpendicular and parallel components of  bidirectional reflectance can be written as,
\begin{equation}\label{7}
 r_{\bot}(i,e,g)=\frac{1}{4\pi}(\frac{\mu_{o}}{\mu+\mu_{o}})\Big\{[\textit{w}p(g)]_{\bot}+\textit{w}[H(\mu_{o})H(\mu)-1]\Big\}
\end{equation}
\begin{equation}\label{8}
r_{\parallel}(i,e,g)=\frac{1}{4\pi}(\frac{\mu_{o}}{\mu+\mu_{o}})\Big\{[\textit{w}p(g)]_{\parallel}+\textit{w}[H(\mu_{o})H(\mu)-1]\Big\}
\end{equation}
where $[{w}p(g)]_\bot$ and $[{w}p(g)]_{\parallel}$ are the radiance scattered once by a particle of the regolith, for incoming radiation with electric vector perpendicular and parallel to the scattering plane respectively.

In laboratory experiment, the value of bidirectional reflectance can be calculated from the relation below (Hapke 1993),
\begin{equation}\label{9}
 r(i,e,g)=\frac{I(i,e,g)}{J}
\end{equation}
where $I(i,e,g)$ and $J$ are respectively the intensities of scattered and incident radiation.

In the laboratory we shall calculate $I(i,e,g)$, by using the Equation (\ref{1a}), after recording scattered intensities at three positions of glan-thomson prism (or the analyser). By using proper normalisation (as well be discussed later), we can calculate the value of $r(i,e,g)$ for our simulation work.

For polarised incident radiation, with electric vector making an angle $45^{0}$ with plane of reflection (or scattering). the total irradiance $J$ can be divided into two equal parts, each of value $\frac{J}{2}$, which are incident on the sample surface with electric vector perpendicular and parallel to the scattering plane. So we can write,
\begin{equation}\label{10}
 J_{\bot}=\frac{J}{2}
\end{equation}
 \begin{equation}\label{11}
 J_{\parallel}=\frac{J}{2}
\end{equation}

where $J_{\bot}$ and $J_{\parallel}$  are the components incident irradiance with electric vector perpendicular and parallel to the scattering plane respectively.

 Substituting values from Equation (\ref{9}), (\ref{10}) in Equation (\ref{7})  and Equation (\ref{9}), (\ref{11}) in Equation (\ref{8}), the perpendicular and parallel components of scattered intensities $I_{\bot}(i,e,g)$ and $I_{\parallel}(i,e,g)$ can be written as follows:
\begin{equation}\label{12}
I_{\bot}(i,e,g)=\frac{J}{2}r_{\bot}(i,e,g)
\end{equation}
\begin{equation}\label{13}
I_{\parallel}(i,e,g)=\frac{J}{2}r_{\parallel}(i,e,g)
\end{equation}

The linear polarisation of the scattered light from the sample surface can be expressed in terms of the stokes parameter $I$, $Q$, $U$ (Stokes 1952),
\begin{equation}\label{14}
P_{model}=\frac{\sqrt{Q^{2}+U^{2}}}{I}
\end{equation}
where
$$I=<E^{2}_{xo}>+<E^{2}_{yo}>$$
$$Q=<E^{2}_{xo}>-<E^{2}_{yo}>$$
$$U=2<E_{xo}E_{yo}cos\delta>$$
$E_{xo}$ and $E_{yo}$ are amplitudes of orthogonal components of light beam; $\delta$ is the phase difference between orthogonal components of light beam and $< >$ indicates  the time average.

Substituting the values of $I$, $Q$ and $U$ in Equation (\ref{14}) and omitting the $< >$ sign for our convenience everywhere, we have
\begin{equation}\label{15}
P_{model}=\frac{\sqrt{(E^{2}_{xo}-E^{2}_{yo})^{2}+4E^{2}_{xo}E^{2}_{yo}cos^{2}\delta}}{E^{2}_{xo}+E^{2}_{yo}}
\end{equation}
The intensities $I_{\bot}$ and $I_{\parallel}$ are actually proportional to  the amplitudes $E_{xo}$ and $E_{yo}$ of orthogonal components of light beam and we may write
\begin{equation}\label{16}
I_{\bot}=KE^{2}_{xo}
\end{equation}
and
\begin{equation}\label{17}
I_{\parallel}=KE^{2}_{yo}
\end{equation}
where $K$ is some arbitrary constant.

Substituting Equation (\ref{16}) and (\ref{17}) in Equation (\ref{15}), we have
\begin{equation}\label{18}
P_{model}=\frac{\sqrt{(I_{\bot}-I_{\parallel})^{2}+4I_{\bot}I_{\parallel}cos^{2}\delta}}{(I_{\bot}+I_{\parallel})}
\end{equation}

Substituting Equation (\ref{12}) and (\ref{13}) in Equation (\ref{18}), we have
\begin{equation}\label{19}
P_{model}=\frac{\sqrt{(r_{\bot}-r_{\parallel})^{2}+4r_{\bot}r_{\parallel}cos^{2}\delta}}{(r_{\bot}+r_{\parallel})}
\end{equation}

In this modeling, we used the following expressions, which can be proved through some logical steps,
\begin{equation}\label{20}
[{w}p(g)]_{\bot}=4{w}_{\bot}p_{\bot}(g)
\end{equation}
\begin{equation}\label{21}
[{w}p(g)]_{\parallel}=4{w}_{\parallel}p_{\parallel}(g)
\end{equation}
\begin{equation}\label{22}
{w}={w}_{\bot}+{w}_{\parallel}
\end{equation}
\begin{equation}\label{23}
p(g)=p_{\bot}(g)+p_{\parallel}(g)
\end{equation}

where ${w}_{\bot}$ and ${w}_{\parallel}$ are the contribution of total single particle albedo $w$ to $I_{\bot}(i,e,g)$ and $I_{\parallel}(i,e,g)$ respectively, $p_{\bot}(g)$ and $p_{\parallel}(g)$ are components single particle scattering phase function $p(g)$ with electric vector perpendicular and parallel to the scattering plane respectively.\\

Substituting Equation (\ref{20}) in Equation (\ref{7}) and Equation (\ref{21}) in Equation (\ref{8}), we have
\begin{equation}\label{24}
r_{\bot}(i,e,g)=\frac{1}{4\pi}(\frac{\mu_{o}}{\mu+\mu_{o}})\Big\{4\textit{w}_{\bot}p_{\bot}(g)+\textit{w}[H(\mu_{o})H(\mu)-1]\Big\}
\end{equation}
\begin{equation}\label{25}
r_{\parallel}(i,e,g)=\frac{1}{4\pi}(\frac{\mu_{o}}{\mu+\mu_{o}})\Big\{4\textit{w}_{\parallel}p(g)_{\parallel}+\textit{w}[H(\mu_{o})H(\mu)-1]\Big\}
\end{equation}

The values of $p_{\bot}(g)$ and $p_{\parallel}(g)$ can be obtained from Mie theory. In appendix, the calculation of ${w}_{\bot}$ and ${w}_{\parallel}$ based on Mie theory is provided, which is done in the present work. The values of $r_{\bot}(i,e,g)$ and $r_{\parallel}(i,e,g)$ can be computed by substituting values of $p_{\bot}(g)$, $p_{\parallel}(g)$, ${w}_{\bot}$ and ${w}_{\parallel}$ from Mie theory and $H(\mu)$,$H(\mu_{0})$ from Equation (\ref{4}) in Equation (\ref{24}) and (\ref{25}) respectively.

From Equation (\ref{24}) and (\ref{25}), the total theoretical bidirectional reflectance $r_{th}(i,e,g)$ can be expressed as:
\begin{equation}\label{26}
r_{th}(i,e,g)=\frac{1}{2}\Big[r_{\bot}(i,e,g)+r_{\parallel}(i,e,g)\Big]
\end{equation}

The above value of $r_{th}(i,e,g)$ as in Equation (\ref{26}), can be identified to be same as $r(i,e,g)$ as expressed in Equation (\ref{3}). This is a theoretically or model calculated value, which will be divided by theoretically calculated reflectance $r(i=45^{o},e=45^{o})$ for normalisation. Thus the normalised reflectance can be expressed as:
\begin{equation}\label{27}
nr_{th}(i,e,g)=\frac{r_{th}(i,e,g)}{r_{th}(45^{o},45^{o},90^{o})}
\end{equation}

This theoretical value will be compared with the observed value as obtained through laboratory experiment. However, this experimentally observed value should be also normalised in the same way.

The experimental value of $r(i,e,g)$ is simply proportional to the intensity value $I$ recorded by our light detector in the laboratory in some arbitrary unit, with the scattering geometry defined by the parameter $(i,e,g)$. The value of $I$ is calculated from $I_{1}$, $I_{2}$ and $I_{3}$ by the relation (\ref{1a}) . The observed normalised reflectance of the scattered light, which is normalised at $(i=45^{o}, e=45^{o}, g=90^{o})$ can be expressed as:

$$nr_{ob}(i,e,g)=\frac{r_{ob}(i,e,g)}{r_{ob}(45^{o},45^{o},90^{o})}\times\frac{\cos i}{\cos e}$$
\begin{equation}\label{28}
=\frac{I(i,e,g)}{I(45^{o},45^{o},90^{o})}\times\frac{\cos i}{\cos e}
\end{equation}

where $(\frac{cosi}{cose})$ takes care of the geometric correction (Deb et al. 2011).

The present work consist of two simulation works, which are as follows:
\begin{enumerate}
\item Photometric simulation :
Here we calculate theoretically the normalised reflectance of scattered light from the sample surface, by using equation (\ref{27}). For experimental values of normalised reflectance, Equation (\ref{28}) is used. We shall compare the two sets of reflectance values.
\item Polarimetric simulation :
Here we calculate theoretically the polarisation of the scattered light from sample surface, by the Equation (\ref{19}) with an unknown parameter $\delta$, which will be used as a free parameter.
\end{enumerate}

The experimental values of polarisation are estimated using Equation (\ref{1b}). Further, here it may be noted that the quantity $(cos\delta)$ as mentioned in Equation (\ref{19}), depends on the phase difference $\delta$ between the two orthogonal components of scattered radiation. The value of $\delta$ depend on some unknown sample parameters which should include roughness, texture and porosity among others. So we prefer to keep it as a free parameter during model fitting.

The modeling routine used for the polarimetric data is a least-squares search over the parameter $\delta$ that minimises $\chi$, which is defined as

\begin{equation}\label{29}
\chi=\frac{\sum(P_{observe}-P_{model})^{2}}{N}
\end{equation}

where $P_{observe}$ and $P_{model}$ are computed from Equation (\ref{1b}) and (\ref{19}) repsectively, N is the number of observations.

\section{Error associated with estimated values of reflectance and  polarisation}

In our experimental work, the intensity values are measured as a function of controlled geometry. We set the angle of incidence and emergence manually and carefully, by noting values on the graduated scale. After that we measure the intensity $I$, on a scale where we have an uncertainty  $\delta I=0.05\times I$. Thus we can assume that, the only uncertainty in our experiment is associated with the measurement of intensity $I$. Thus from Equation (\ref{28}), it can be shown that the uncertainty in normalised reflectance can be expressed as:
\begin{equation}
\delta nr=\sqrt{2}\frac{\delta I}{I}\times nr
\end{equation}

Since the nr value can be as large as 1 and $\frac{\delta I}{I}\sim0.05$, we can conclude that the maximum uncertainty in measurement of nr can be $\delta nr=0.07$.

The uncertainity in polarimetric measurement can be calculated from the relation given below (Sen et al. 1900):
\begin{equation}\label{2}
\delta P \approx \big(\frac{2+P^{2}}{3}\big)^{\frac{1}{2}}(\frac{\delta I}{I})
\end{equation}

Since $\frac{\delta I}{I}\sim0.05$, as discussed above and the maximum value of $P$ as observed in our experiment is of the order of 10\%, we can safely take the upper limit on $\delta P$  as $\delta P< 4.18\%$

\setlength{\abovecaptionskip}{50pt}
\begin{figure*}[!htbp]
\centering
\includegraphics[width=15cm,height=10cm]{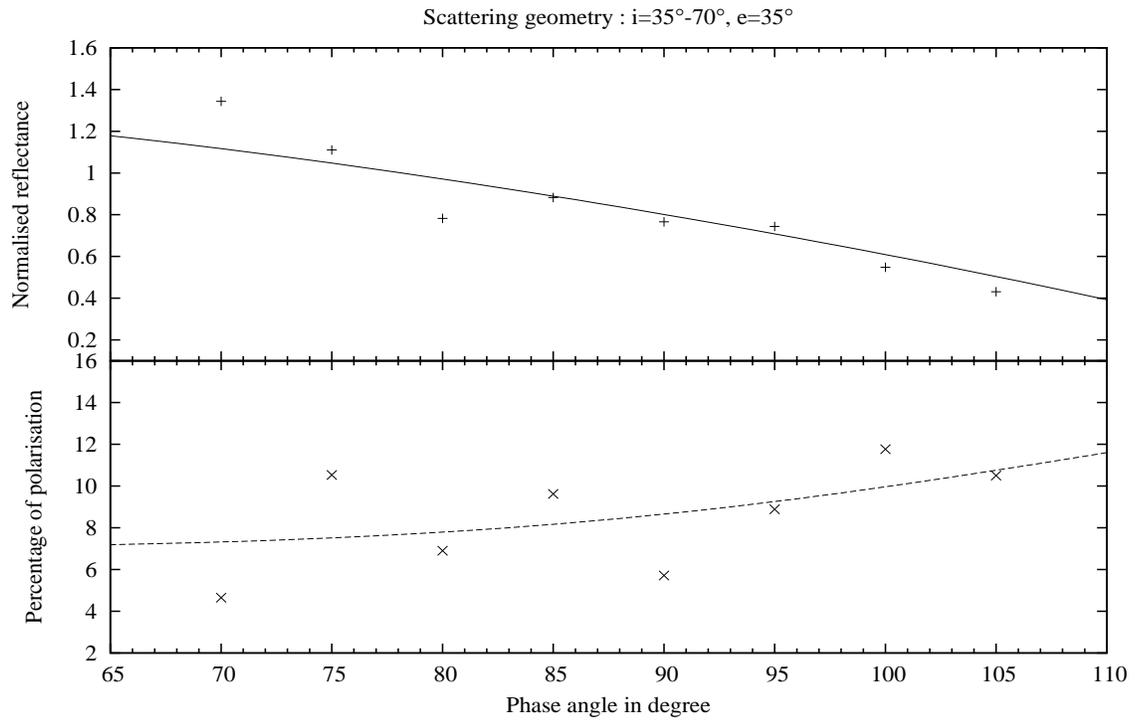}\\
\caption{Normalised reflectance and percentage of polarisation of light scattered from a rough surface containing 0.3$\mu$m sized Alumina powder are shown as a function of phase angle. The + and $\times$ sign represent experimentally observed data of normalised reflectance and percentage of polarisation respectively obtained from the ellipsometric measurement. The solid and dashed lines represent the theoretical curve of normalised reflectance and percentage of polarisation respectively obtained by using Hapke model with Mie theory.   However, the polarisation curve was fitted to the observed data, by using a fitting procedure as discussed in section 4. The best fitted curve for polarisation is obtained at $\delta=86^{0}$.}
\end{figure*}
\vspace{5cm}

\setlength{\abovecaptionskip}{50pt}
\begin{figure*}[!htbp]
\centering
\includegraphics[width=15cm,height=10cm]{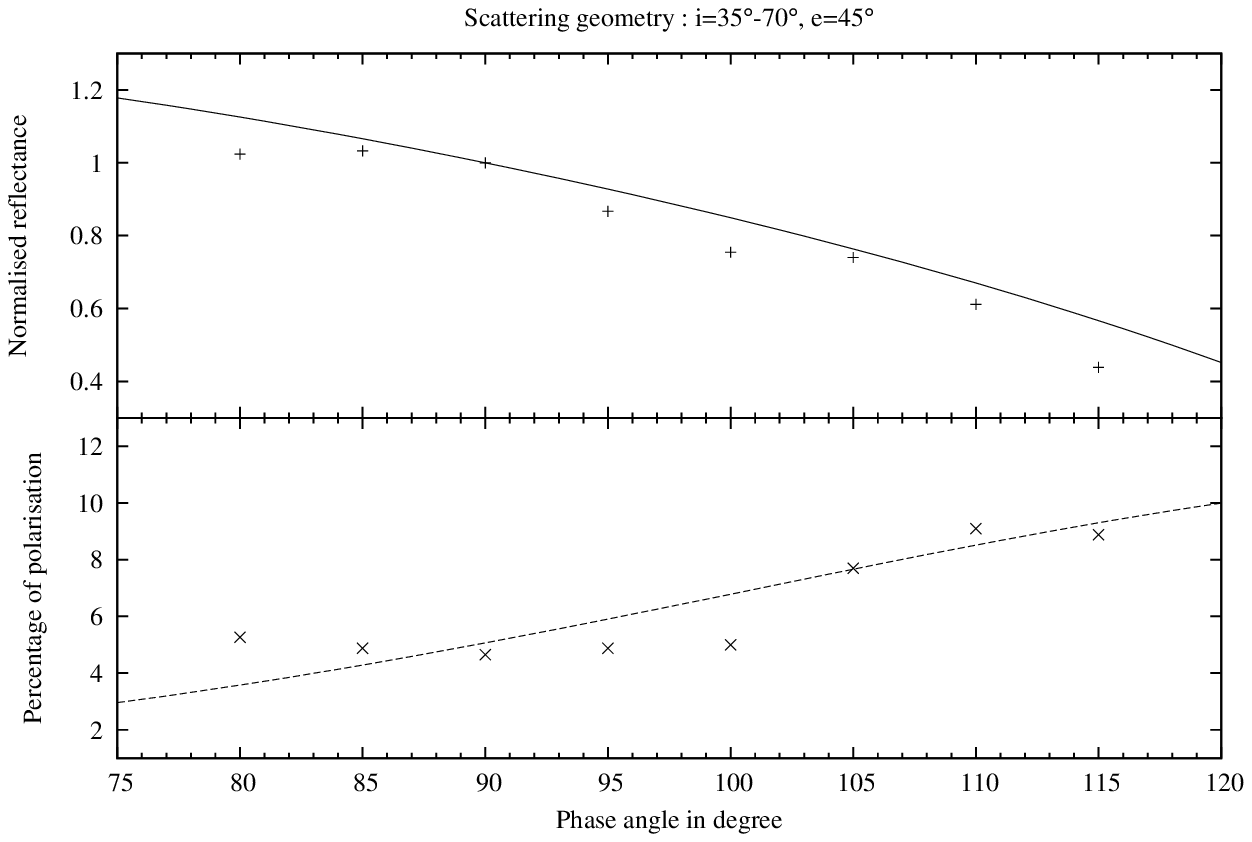}\\
\caption{Normalised reflectance and percentage of polarisation of light scattered from a rough surface containing 0.3$\mu$m sized Alumina powder are shown as a function of phase angle. The + and $\times$ sign represent experimentally observed data of normalised reflectance and percentage of polarisation respectively obtained from the ellipsometric measurement. The solid and dashed lines represent the theoretical curve of normalised reflectance and percentage of polarisation respectively obtained by using Hapke model with Mie theory.   However, the polarisation curve was fitted to the observed data, by using a fitting procedure as discussed in section 4. The best fitted curve for polarisation is obtained at $\delta=89.5^{0}$.}
\end{figure*}
\vspace{5cm}

\setlength{\abovecaptionskip}{50pt}
\begin{figure*}[!htbp]
\centering
\includegraphics[width=15cm,height=10cm]{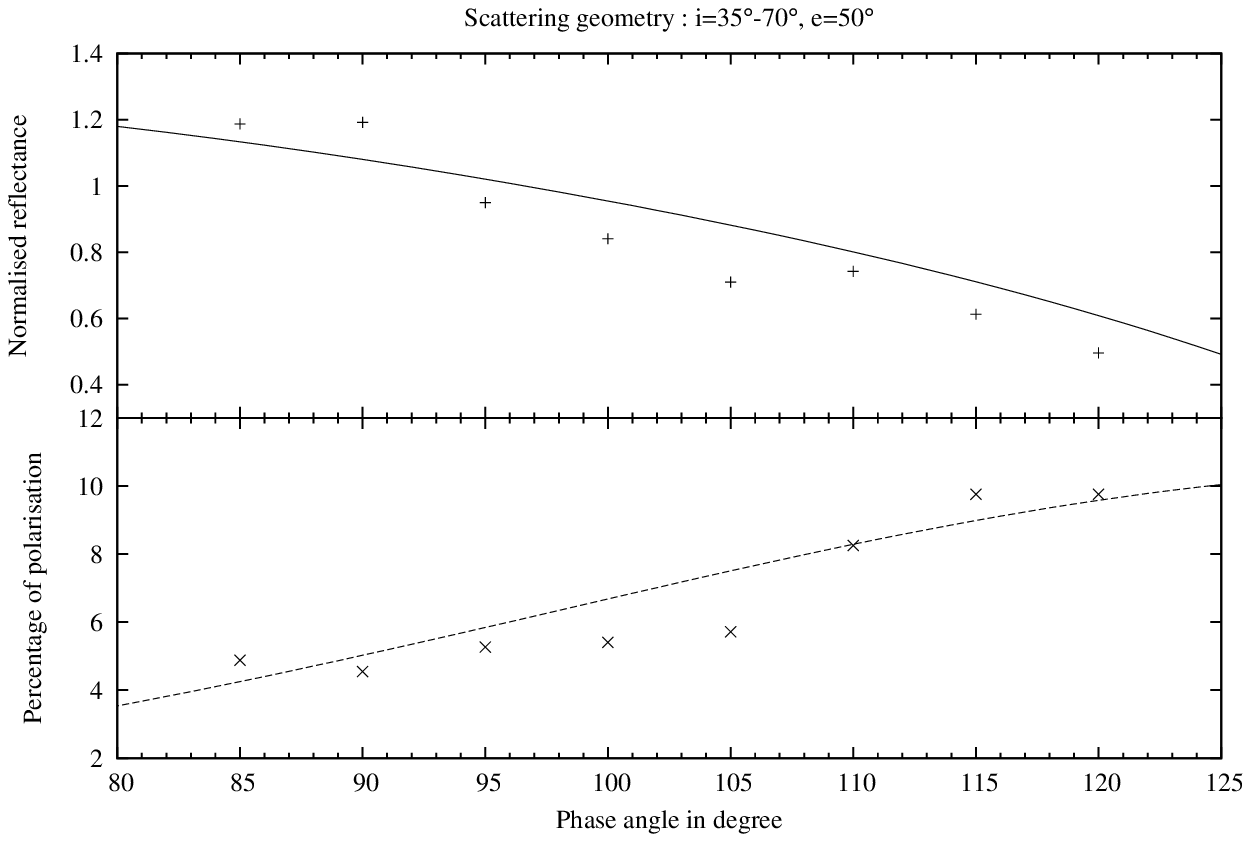}\\
\caption{Normalised reflectance and percentage of polarisation of light scattered from a rough surface containing 0.3$\mu$m sized Alumina powder are shown as a function of phase angle. The + and $\times$ sign represent experimentally observed data of normalised reflectance and percentage of polarisation respectively obtained from the ellipsometric measurement. The solid and dashed lines represent the theoretical curve of normalised reflectance and percentage of polarisation respectively obtained by using Hapke model with Mie theory.   However, the polarisation curve was fitted to the observed data, by using a fitting procedure as discussed in section 4. The best fitted curve for polarisation is obtained at $\delta=90^{0}$.}
\end{figure*}
\vspace{5cm}

\setlength{\abovecaptionskip}{50pt}
\begin{figure*}[!htbp]
\centering
\includegraphics[width=15cm,height=10cm]{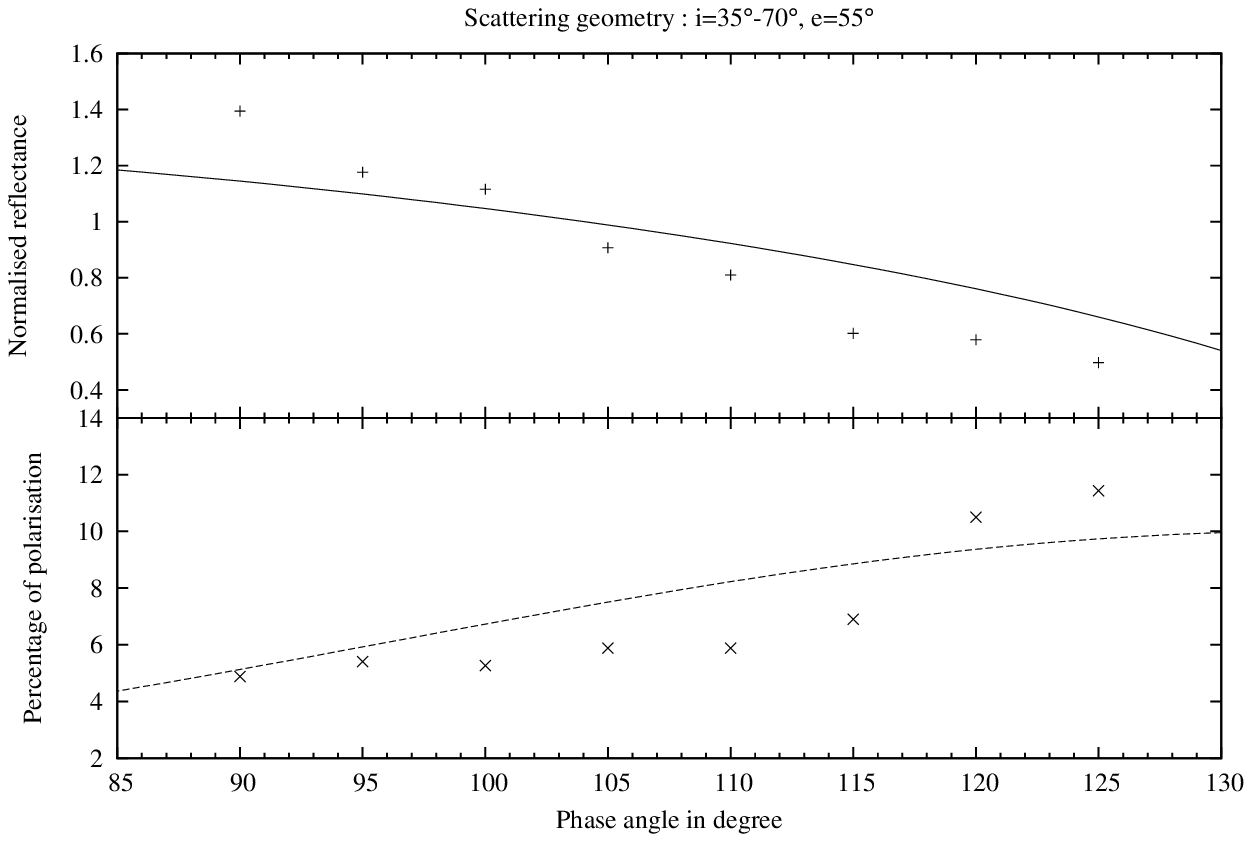}\\
\caption{Normalised reflectance and percentage of polarisation of light scattered from a rough surface containing 0.3$\mu$m sized Alumina powder are shown as a function of phase angle. The + and $\times$ sign represent experimentally observed data of normalised reflectance and percentage of polarisation respectively obtained from the ellipsometric measurement. The solid and dashed lines represent the theoretical curve of normalised reflectance and percentage of polarisation respectively obtained by using Hapke model with Mie theory.   However, the polarisation curve was fitted to the observed data, by using a fitting procedure as discussed in section 4. The best fitted curve for polarisation is obtained at $\delta=90^{0}$.}
\end{figure*}
\vspace{5cm}

\setlength{\abovecaptionskip}{50pt}
\begin{figure*}[!htbp]
\centering
\includegraphics[width=15cm,height=10cm]{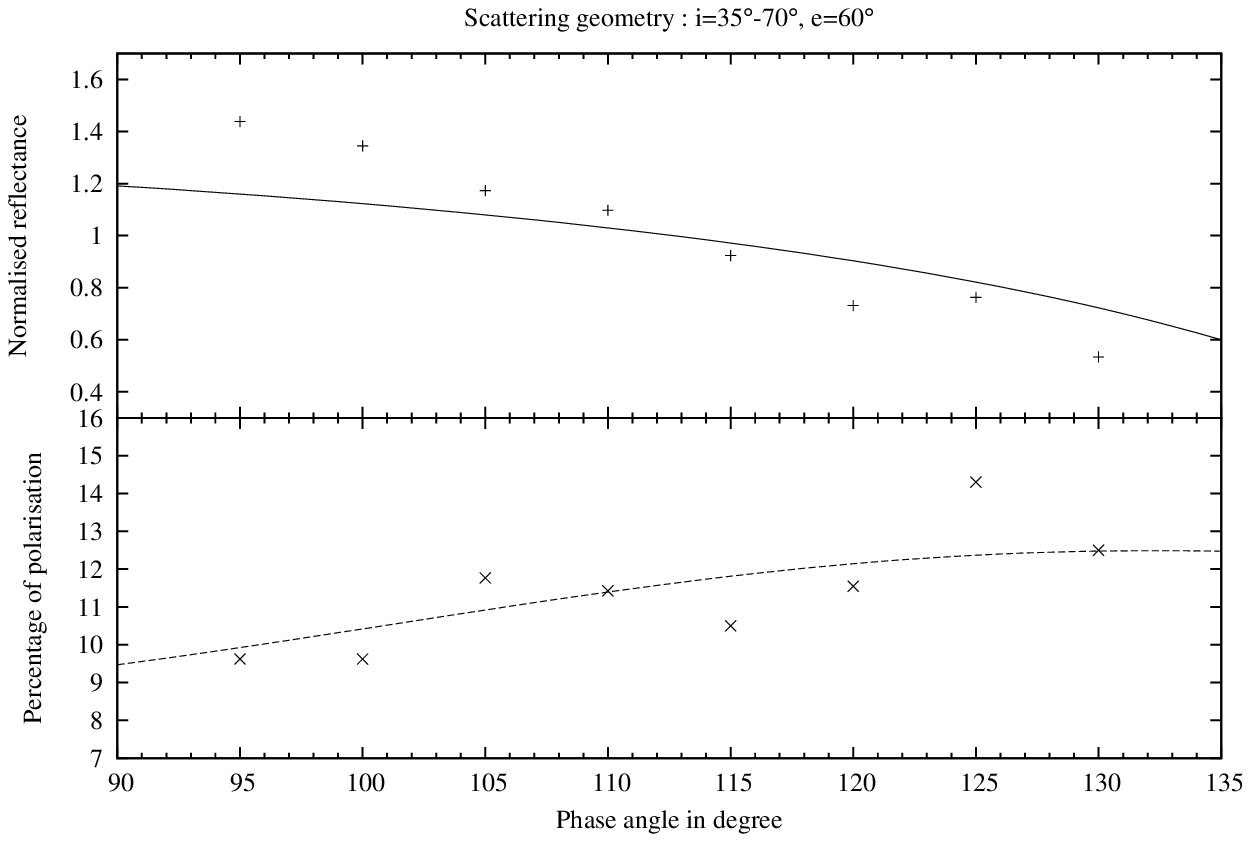}\\
\caption{Normalised reflectance and percentage of polarisation of light scattered from a rough surface containing 0.3$\mu$m sized Alumina powder are shown as a function of phase angle. The + and $\times$ sign represent experimentally observed data of normalised reflectance and percentage of polarisation respectively obtained from the ellipsometric measurement. The solid and dashed lines represent the theoretical curve of normalised reflectance and percentage of polarisation respectively obtained by using Hapke model with Mie theory.   However, the polarisation curve was fitted to the observed data, by using a fitting procedure as discussed in section 4. The best fitted curve for polarisation is obtained at $\delta=85.5^{0}$.}
\end{figure*}
\vspace{5cm}

\setlength{\abovecaptionskip}{50pt}
\begin{figure*}[!htbp]
\centering
\includegraphics[width=15cm,height=10cm]{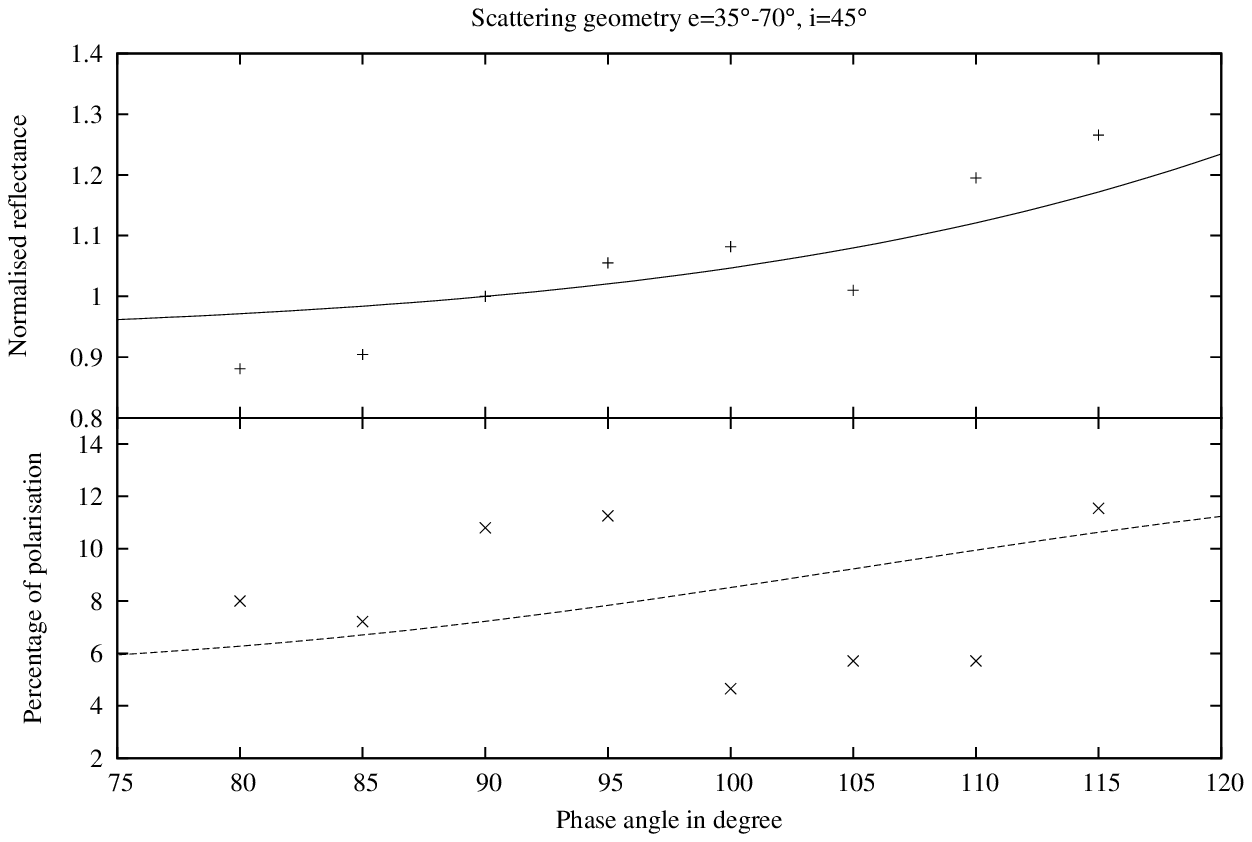}\\
\caption{Normalised reflectance and percentage of polarisation of light scattered from a rough surface containing 0.3$\mu$m sized Alumina powder are shown as a function of phase angle. The + and $\times$ sign represent experimentally observed data of normalised reflectance and percentage of polarisation respectively obtained from the ellipsometric measurement. The solid and dashed lines represent the theoretical curve of normalised reflectance and percentage of polarisation respectively obtained by using Hapke model with Mie theory.   However, the polarisation curve was fitted to the observed data, by using a fitting procedure as discussed in section 4. The best fitted curve for polarisation is obtained at $\delta=87^{0}$.}
\end{figure*}
\vspace{5cm}

\setlength{\abovecaptionskip}{50pt}
\begin{figure*}[!htbp]
\centering
\includegraphics[width=15cm,height=10cm]{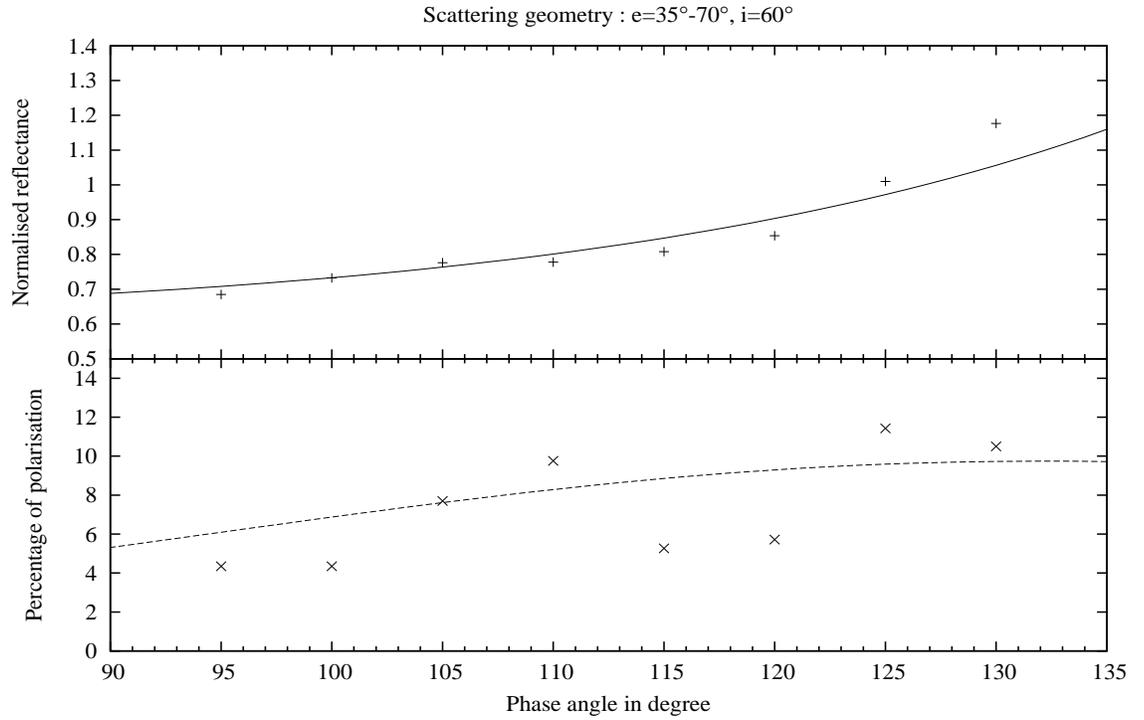}\\
\caption{Normalised reflectance and percentage of polarisation of light scattered from a rough surface containing 0.3$\mu$m sized Alumina powder are shown as a function of phase angle. The + and $\times$ sign represent experimentally observed data of normalised reflectance and percentage of polarisation respectively obtained from the ellipsometric measurement. The solid and dashed lines represent the theoretical curve of normalised reflectance and percentage of polarisation respectively obtained by using Hapke model with Mie theory.   However, the polarisation curve was fitted to the observed data, by using a fitting procedure as discussed in section 4. The best fitted curve for polarisation is obtained at $\delta=90^{0}$.}
\end{figure*}
\vspace{5cm}

\section{Result and discussion}
The data points in the figure 2-8 are computed from the data generated by Ellipsometer. Each figure consists of two panels. The upper panel represents the variation of normalised reflectance of the scattered light as a function of phase angle for 0.3$\mu$m sized powder of $Al_{2}O_{3}$. The model curve in upper panels is obtained by using Equation (\ref{27}).  The lower panel represents the variation of percentage of polarisation $P$ of the scattered light as a function of phase angle, which is calculated from same data, under identical conditions. The model curve in lower panel is obtained by using Equation (\ref{19}). Here we noted that for reflectance data, no curve fitting algorithm was used. The experimental data and theoretical curve were plotted as they are. However, for polarisation curve, a $\chi^{2}$  minimisation technique, for fitting the data was applied with phase difference $\delta$ as free parameter as discussed below.

The refractive index of $Al_{2}O_{3}$ is n=1.766 at 632.8$\mu$m (Gervais1991). In the work of Deb et al. 2011, the bidirectional reflectance data for 0.3$\mu$m $Al_{2}O_{3}$ is studied by combining Hapke model with Mie theory. The best fitted value of absorption coefficient k=0.000009 at 632.8$\mu$m was obtained by Deb et al. 2011, for matching observed data with Hapke model. In the present work, the complex optical constant (n=1.766,k=0.000009) at 632.8$\mu$m for $Al_{2}O_{3}$ (as obtained by Deb et al. 2011)is used to compute $r_{\bot}(i,e,g)$ and $r_{\parallel}(i,e,g)$ from Equation (\ref{24}) and (\ref{25}). These values of reflectance are used to compute the theoretical normalised reflectance from Equation (\ref{27}) to model the experimental normalised reflectance data for different scattering geometries. So no free parameter is used for fitting of the modeling for normaliased reflectance.

 The values of  $r_{\bot}(i,e,g)$ and $r_{\parallel}(i,e,g)$, which were computed previously, are now used in Equation (\ref{19}) to theoretically calculate percentage of polarisation. In order to fit the experimental data to this theoretical polarisation curve from the model (as in Equation (\ref{18})), the phase difference $\delta$  is used as a free parameter. Towards this fitting procedure, the   $\chi^{2}$- minimisation technique was used, as was previously done by Deb et al. 2011, for fitting reflectance data. The best fitted values of $\delta$ for different scattering geometries are lying in the range $[85.5^{0}-90^{0}]$, which are listed in table 1.

In both types of modeling viz. for normalised reflectance $nr$ and percentage of polarisation $P$; the grain size, optical constant and monochromatic wavelength of incident radiation are taken as input parameters.

\vspace{5mm}
\setlength{\abovecaptionskip}{20pt}
\begin{table}[!htbp]
\begin{tabular}{ |c|c|c| }
 \hline
  Angle of incidence i& Angle of emergence e & Best fitted phase difference $\delta$\\
   \hline
 $(35^{0}-70^{0})$ & $35^{0}$ & $86^{0}$ \\
 $(35^{0}-70^{0})$ & $45^{0}$ & $89.5^{0}$ \\
 $(35^{0}-70^{0})$ & $50^{0}$ & $90^{0}$ \\
 $(35^{0}-70^{0})$ & $55^{0}$ & $90^{0}$ \\
 $(35^{0}-70^{0})$ & $60^{0}$ & $85.5^{0}$ \\
 $45^{0}$ & $(35^{0}-70^{0})$ & $87^{0}$ \\
 $60^{0}$ & $(35^{0}-70^{0})$ & $90^{0}$ \\
  \hline
\end{tabular}
\caption{Table of best fitted values of free parameter $\delta$}
\label{table:1}
\end{table}

The agreement between the normalised reflectance data with the corresponding model curve is obtained for 0.3$\mu$m $Al_{2}O_{3}$, has been shown in upper panels of figure 2-8.  This agreement proves the correctness of measured intensities, from which the normalised reflectance is computed by using Equation (\ref{28}), in other words, using Hapke's model combined with Mie Theory. From the same set of intensities, the percentage of polarisation is computed by using Equation (\ref{1b}). Whereas, the theoretical values of polarisation has been estimated by  using Equation (\ref{19}), which has been derived in the present work based on our proposed model. For modeling this polarimetric data, however in literature no suitable procedure is available till today, which can successfully explain the data in terms of the physical properties of constituent grains of rough surface from which the scattering is taking place. For this reason, in the present work a new polarimetric model is developed based on Mie theory and Hapke model, to the explain polarimetric data obtained for 0.3$\mu$m $Al_{2}O_{3}$. The single particle physical properties viz. grain size, complex optical constant of grains and size parameter are introduced in present modeling by Mie theory. The agreement between the percentage of polarisation  and corresponding model curve based on new polarimetric model, is obtained by taking phase difference $\delta$ as a free fitting parameter.  In lower panels of figure 2-8, this comparison has been shown. This agreement proves the correctness of the model developed in the present work. It should be also noted that, though phase difference ($\delta $) has been used as a free parameter, but for the seven independent data sets the value of $\delta$ has been found to lie within a narrow range viz. $85^{0}- 90^{0}$. This fact further strengthens the consistency of our model. In future, this model can be tested for other samples to know further the range of applicability of our model. Finally, this model can be used to predict the polarization caused due to rough surface scattering. Alternately, form the observed polarization data from rough surface, one would be able to infer about the properties of constituent grains, by using this model.

\section*{Acknowledgements}
This work is supported and financed by UGC-SAP grant ( 530/12/DRS/2010 (SAP-1) dtd. 01 Mar 2013). The authors are thankful to Amritaksha Kar,  Dept. of Physics, Assam University, Silchar for his immense support and valuable suggestions in performing the laboratory work.The authors are also thankful to Dr. A. Deshmukhya, Dept. of Physics, Assam University, Silchar for her moral support and encouragement during this work.
\vspace{5cm}

\renewcommand{\theequation}{A-\arabic{equation}}
  % redefine the command that creates the equation no.
\setcounter{equation}{0}  % reset counter
\section*{APPENDIX}
In Mie theory, when an electromagnetic radiation of wavelength $\lambda$ interacts with a spherical particle of radius $a$ with complex refractive index $m$, the total scattering efficiency can be expressed as (Van de Hulst 1981):
\begin{equation}\label{A19}
Q_{sca}=\frac{1}{x^{2}}\int_{0}^{\pi}[i_{1}(\theta)+i_{2}(\theta)]\sin\theta d\theta
\end{equation}

where $\theta$ is the scattering angle, $x=\frac{2\pi a}{\lambda}$ is the size parameter; $i_{1}(\theta)$ and $i_{2}(\theta)$ are intensity functions associated with the electric vector perpendicular and parallel to the scattering plane respectively.  

From Equation (\ref{A19}), the perpendicular and parallel components of total scattering efficiency $Q_{sca}$ can be expressed as:
\begin{equation}\label{A20}
Q_{sca\perp}=\frac{1}{x^{2}}\int_{0}^{\pi}i_{1}(\theta)\sin\theta d\theta
\end{equation}
\begin{equation}\label{A21}
Q_{sca\parallel}=\frac{1}{x^{2}}\int_{0}^{\pi}i_{2}(\theta)\sin\theta d\theta
\end{equation}
The expressions of intensity functions $i_{1}(\theta)$ and $i_{2}(\theta)$ can be expressed as (Van de Hulst 1981):
\begin{equation}\label{A22}
i_{1}(\theta)=|S_{1}(\theta)|^{2}
\end{equation}
\begin{equation}\label{A23}
i_{2}(\theta)=|S_{2}(\theta)|^{2}
\end{equation}
where $S_{1}(\theta)$ and $S_{2}(\theta)$ are amplitude functions associated with $i_{1}(\theta)$ and $i_{2}(\theta)$ respectively.

The expressions of amplitude functions $S_{1}(\theta)$ and $S_{2}(\theta)$ (Van de Hulst 1981) can be expressed as,
\begin{equation}\label{A24}
S_{1}(\theta)=\sum_{n=1}^{\infty}\frac{2n+1}{n(n+1)}[a_{n}\pi_{n}(\cos \theta)+b_{n}\tau_{n}(\cos \theta)]
\end{equation}
\begin{equation}\label{A25}
S_{2}(\theta)=\sum_{n=1}^{\infty}\frac{2n+1}{n(n+1)}[b_{n}\pi_{n}(\cos \theta)+a_{n}\tau_{n}(\cos \theta)]
\end{equation}
where $a_{n}$ and $b_{n}$ are function of size parameter $x$ and complex refractive index $m$ of the sphere,
\begin{equation}\label{A26}
\pi_{n}(\cos\theta)=\frac{1}{\sin \theta}P^{1}_{n}(\cos \theta)
\end{equation}
\begin{equation}\label{A27}
\tau_{n}(\cos\theta)=\frac{d}{d\theta}P^{1}_{n}(\cos \theta)
\end{equation}
and $P^{1}_{n}$ is associated Legendre function of first kind of order 1 and degree n.

Substituting Equation (\ref{A24}) in Equation (\ref{A22}), $i_{1}(\theta)$ can be written as,
$$
i_{1}(\theta)=|S_{1}(\theta)|^{2}=\sum_{n=1}^{\infty}\frac{2n+1}{n(n+1)}[a_{n}\pi_{n}+b_{n}\tau_{n}]\times\sum_{m=1}^{\infty}\frac{2m+1}{m(m+1)}[a_{m}\pi_{m}+b_{m}\tau_{m}]
$$
\begin{equation}\label{A28}
\Rightarrow i_{1}(\theta)=\sum_{n=1}^{\infty}\sum_{m=1}^{\infty}\frac{2n+1}{n(n+1)}\frac{2m+1}{m(m+1)}(a_{n}a_{m}\pi_{n}\pi_{m}+a_{n}b_{m}\pi_{n}\tau_{m}+a_{m}b_{n}\pi_{m}\tau_{n}+b_{n}b_{m}\tau_{n}\tau_{m})
\end{equation}

Also substituting Equation (\ref{A25}) in Equation (\ref{A23}), $i_{2}(\theta)$ can be written as,

$$
i_{2}(\theta)=|S_{2}(\theta)|^{2}=\sum_{i=1}^{\infty}\frac{2i+1}{i(i+1)}[b_{i}\pi_{i}+a_{i}\tau_{i}]\times\sum_{j=1}^{\infty}\frac{2j+1}{j(j+1)}[b_{j}\pi_{j}+a_{j}\tau_{j}]
$$
\begin{equation}\label{A29}
\Rightarrow i_{2}(\theta)=\sum_{i=1}^{\infty}\sum_{j=1}^{\infty}\frac{2i+1}{i(i+1)}\frac{2j+1}{j(j+1)}(b_{i}b_{j}\pi_{i}\pi_{j}+b_{i}a_{j}\pi_{i}\tau_{j}+a_{i}b_{j}\tau_{i}\pi_{j}+a_{i}a_{j}\tau_{i}\tau_{j})
\end{equation}

Now combining Equation (\ref{A28}) and Equation (\ref{A20}), the perpendicular component of scattering efficiency $Q_{sca\perp}$ can be written as,
$$
Q_{sca\perp}=\frac{1}{x^{2}}\int_{0}^{\pi}\sum_{n=1}^{\infty}\sum_{m=1}^{\infty}\frac{2n+1}{n(n+1)}\frac{2m+1}{m(m+1)}\Bigg(a_{n}a_{m}\pi_{n}\pi_{m}+a_{n}b_{m}\pi_{n}\tau_{m}+a_{m}b_{n}\pi_{m}\tau_{n}+b_{n}b_{m}\tau_{n}\tau_{m}\Bigg)\sin\theta d\theta
$$
$$
=\frac{1}{x^{2}}\sum_{n=1}^{\infty}\sum_{m=1}^{\infty}\frac{2n+1}{n(n+1)}\frac{2m+1}{m(m+1)}\Bigg(a_{n}a_{m}\int_{0}^{\pi}\pi_{n}\pi_{m}\sin\theta d\theta+a_{n}b_{m}\int_{0}^{\pi}\pi_{n}\tau_{m}\sin\theta d\theta
$$
\begin{equation}\label{A30}
+a_{m}b_{n}\int_{0}^{\pi}\pi_{m}\tau_{n}\sin\theta d\theta+b_{n}b_{m}\int_{0}^{\pi}\tau_{n}\tau_{m}\sin\theta d\theta\Bigg)
\end{equation}

Also combining Equation (\ref{A29}) and Equation (\ref{A21}), the parallel component of scattering efficiency $Q_{sca\parallel}$ can be written as,
$$
Q_{sca\parallel}=\frac{1}{x^{2}}\int_{0}^{\pi}\sum_{i=1}^{\infty}\sum_{j=1}^{\infty}\frac{2i+1}{i(i+1)}\frac{2j+1}{j(j+1)}\Bigg(b_{i}b_{j}\pi_{i}\pi_{j}+b_{i}a_{j}\pi_{i}\tau_{j}+a_{i}b_{j}\tau_{i}\pi_{j}+a_{i}a_{j}\tau_{i}\tau_{j}\Bigg)\sin\theta d\theta
$$
$$
=\frac{1}{x^{2}}\sum_{i=1}^{\infty}\sum_{j=1}^{\infty}\frac{2i+1}{i(i+1)}\frac{2j+1}{j(j+1)}\Bigg(b_{i}b_{j}\int_{0}^{\pi}\pi_{i}\pi_{j}\sin\theta d\theta+b_{i}a_{j}\int_{0}^{\pi}\pi_{i}\tau_{j}\sin\theta d\theta
$$
\begin{equation}\label{A31}
+a_{i}b_{j}\int_{0}^{\pi}\tau_{i}\pi_{j}\sin\theta d\theta+a_{i}a_{j}\int_{0}^{\pi}\tau_{i}\tau_{j}\sin\theta d\theta\Bigg)
\end{equation}

From Equation (\ref{A30}) and (\ref{A31}), it is found that the following four integrations have to be performed in order to calculate the values of $Q_{sca\perp}$ and $Q_{sca\parallel}$,
\begin{equation}\label{A32}
\int_{0}^{\pi}\pi_{p}\pi_{q}\sin\theta d\theta
\end{equation}
\begin{equation}\label{A33}
\int_{0}^{\pi}\pi_{p}\tau_{q}\sin\theta d\theta
\end{equation}
\begin{equation}\label{A34}
\int_{0}^{\pi}\pi_{q}\tau_{p}\sin\theta d\theta
\end{equation}
\begin{equation}\label{A35}
\int_{0}^{\pi}\tau_{p}\tau_{q}\sin\theta d\theta
\end{equation}

where the index $p$ represents $n$ (in Equation (\ref{A30})) and $i$ (in Equation (\ref{A31})). Similarly the index $q$ represents $m$ (in Equation (\ref{A30})) and $j$ (in Equation (\ref{A31})).

Using Equation (\ref{A26}) in the expression (\ref{A32}), we have
$$
\int_{0}^{\pi}\pi_{p}\pi_{q}\sin\theta d\theta=\int_{0}^{\pi}\frac{1}{\sin \theta}P^{1}_{p}(\cos \theta)\frac{1}{\sin \theta}P^{1}_{q}(\cos \theta)\sin\theta d\theta
$$
\begin{equation}\label{A1}
=\int_{0}^{\pi}\frac{1}{\sin^{2}\theta}P^{1}_{p}(\cos \theta)P^{1}_{q}(\cos \theta)\sin\theta d\theta
\end{equation}

Let $x=cos\theta$
$\Rightarrow dx=-sin\theta d\theta$

When $\theta=\pi$, $x=-1$ and $\theta=0$, $x=1$

Using the above substitutions in Equation (\ref{A1}), we have

\begin{equation}\label{A36}
\int_{0}^{\pi}\pi_{p}\pi_{q}\sin\theta d\theta=\int_{-1}^{1}\frac{P^{1}_{p}(x)P^{1}_{q}(x)}{1-x^{2}}dx
\end{equation}

For $p=q$, Equation (\ref{A36}) reduces to,
\begin{equation}\label{A37}
\int_{0}^{\pi}\pi^{2}_{p}\sin\theta d\theta=\int_{-1}^{1}\frac{[P^{1}_{p}(x)]^{2}}{1-x^{2}}dx
\end{equation}

 From Riley et al. 2006, we have
\begin{equation}\label{A38}
\int_{-1}^{1}\frac{[P^{q}_{p}(x)]^{2}}{1-x^{2}}dx=\frac{(p+q)!}{q(p-q)!}
\end{equation}

For $q$=1, Equation (\ref{A38}) reduces to,
$$\int_{-1}^{1}\frac{[P^{1}_{p}(x)]^{2}}{1-x^{2}}dx=\frac{(p+1)!}{(p-1)!}$$
$$=\frac{(p+1)p(p-1)!}{(p-1)!}$$
\begin{equation}\label{A39}
=p(p+1)
\end{equation}

Using Equation (\ref{A39}) in Equation (\ref{A37}), we have
\begin{equation}\label{A40}
\int_{0}^{\pi}\pi^{2}_{p}\sin\theta d\theta=p(p+1)
\end{equation}

The relation between $\pi_{p}$ and $\tau_{q}$, from Bohren and Huffman 1983 is reproduced below,
\begin{equation}\label{A41}
\int_{0}^{\pi}(\pi_{p}\pi_{q}+\tau_{p}\tau_{q})\sin{\theta} d\theta=\delta_{pq}\frac{2p^{2}(p+1)^{2}}{2p+1}
\end{equation}

where $\delta_{pq}$= kronecker delta

And also we have,
$$\int_{0}^{\pi}(\pi_{p}\tau_{q}+\pi_{q}\tau_{p})\sin{\theta} d\theta=\int_{0}^{\pi}(\frac{1}{\sin\theta}P^{1}_{p}\frac{dP^{1}_{q}}{d\theta}+\frac{1}{\sin\theta}P^{1}_{q}\frac{dP^{1}_{p}}{d\theta})\sin\theta d\theta$$
$$=\int_{0}^{\pi}(P^{1}_{p}\frac{dP^{1}_{q}}{d\theta}+P^{1}_{q}\frac{dP^{1}_{p}}{d\theta}) d\theta$$
$$=\int_{0}^{\pi}\frac{d}{d\theta}(P^{1}_{p}P^{1}_{q})d\theta$$
$$=[P^{1}_{p}P^{1}_{q}]^{\pi}_{0}$$
$$=0$$

\begin{equation}\label{A42}
\Rightarrow\int_{0}^{\pi}\pi_{p}\tau_{q}\sin{\theta} d\theta=-\int_{0}^{\pi}\pi_{q}\tau_{p}\sin{\theta} d\theta
\end{equation}

For $p=q$, the Equation (\ref{A41}) reduces to,
$$\int_{0}^{\pi}(\pi^{2}_{p}+\tau^{2}_{p})\sin{\theta} d\theta=\frac{2p^{2}(p+1)}{2p+1}$$
\begin{equation}\label{A43}
\Rightarrow\int_{0}^{\pi}\tau^{2}_{p}\sin{\theta} d\theta=\frac{2p^{2}(p+1)^{2}}{2p+1}-\int_{0}^{\pi}\pi^{2}_{p}\sin{\theta} d\theta
\end{equation}

Combining Equation (\ref{A40}) and Equation (\ref{A43}), we have

$$\int_{0}^{\pi}\tau^{2}_{p}\sin{\theta} d\theta=\frac{2p^{2}(p+1)^{2}}{2p+1}-p(p+1)$$
$$=p(p+1)[\frac{2p(p+1)-2p-1}{2p+1}]$$
$$=p(p+1)[\frac{2p^{2}+2p-2p-1}{2p+1}]$$

\begin{equation}\label{A44}
\Rightarrow\int_{0}^{\pi}\tau^{2}_{p}\sin{\theta} d\theta=\frac{p(p+1)(2p^{2}-1)}{2p+1}
\end{equation}

Combining Equation (\ref{A42}) and Equation (\ref{A30}), we have
$$
Q_{sca\perp}=\frac{1}{x^{2}}\sum_{n=1}^{\infty}\sum_{m=1}^{\infty}\frac{2n+1}{n(n+1)}\frac{2m+1}{m(m+1)} \Bigg(a_{n}a_{m}\int_{0}^{\pi}\pi_{n}\pi_{m}\sin\theta d\theta+a_{n}b_{m}\int_{0}^{\pi}\pi_{n}\tau_{m}\sin\theta d\theta
$$
$$
-a_{m}b_{n}\int_{0}^{\pi}\pi_{n}\tau_{m}\sin\theta d\theta+b_{n}b_{m}\int_{0}^{\pi}\tau_{n}\tau_{m}\sin\theta d\theta\Bigg)
$$
$$
\Rightarrow Q_{sca\perp}=\frac{1}{x^{2}}\sum_{n=1}^{\infty}\sum_{m=1}^{\infty}\frac{2n+1}{n(n+1)}\frac{2m+1}{m(m+1)}\Bigg[a_{n}a_{m}\int_{0}^{\pi}\pi_{n}\pi_{m}\sin\theta d\theta+(a_{n}b_{m}-a_{m}b_{n})\int_{0}^{\pi}\pi_{n}\tau_{m}\sin\theta d\theta
$$
\begin{equation}\label{A45}
+b_{n}b_{m}\int_{0}^{\pi}\tau_{n}\tau_{m}\sin\theta d\theta\Bigg]
\end{equation}

For $n=m$, Equation (\ref{A45}) reduces to,

\begin{equation}\label{A46}
Q_{sca\perp}=\frac{1}{x^{2}}\sum_{n=1}^{\infty}\Bigg[\frac{2n+1}{n(n+1)}\Bigg]^{2}\Bigg[\mid a_{n}\mid^{2}\int_{0}^{\pi}\pi^{2}_{n}\sin\theta d\theta+\mid b_{n}\mid^{2}\int_{0}^{\tau}\tau^{2}_{n}\sin\theta d\theta\Bigg]
\end{equation}
Combing Equation (\ref{A42}) and Equation (\ref{A31}), we have
$$
Q_{sca\parallel}=\frac{1}{x^{2}}\sum_{i=1}^{\infty}\sum_{j=1}^{\infty}\frac{2i+1}{i(i+1)}\frac{2j+1}{j(j+1)}\Bigg(b_{i}b_{j}\int_{0}^{\pi}\pi_{i}\pi_{j}\sin\theta d\theta-b_{i}a_{j}\int_{0}^{\pi}\tau_{i}\pi_{j}\sin\theta d\theta
$$
$$+a_{i}b_{j}\int_{0}^{\pi}\tau_{i}\pi_{j}\sin\theta d\theta+a_{i}a_{j}\int_{0}^{\pi}\tau_{i}\tau_{j}\sin\theta d\theta\Bigg)$$

\begin{equation}\label{A47}
\begin{split}
\Rightarrow Q_{sca\parallel}=\frac{1}{x^{2}}\sum_{i=1}^{\infty}\sum_{j=1}^{\infty}\frac{2i+1}{i(i+1)}\frac{2j+1}{j(j+1)}\Bigg[b_{i}b_{j}\int_{0}^{\pi}\pi_{i}\pi_{j}\sin\theta d\theta \\
+(a_{i}b_{j}-b_{i}a_{j})\int_{0}^{\pi}\tau_{i}\pi_{j}\sin\theta d\theta+a_{i}a_{j}\int_{0}^{\pi}\tau_{i}\tau_{j}\sin\theta d\theta\Bigg]
\end{split}
\end{equation}

For $i=j$, Equation (\ref{A47}) reduces to,

\begin{equation}\label{A48}
Q_{sca\parallel}=\frac{1}{x^{2}}\sum_{i=1}^{\infty}\Bigg[\frac{2i+1}{i(i+1)}\Bigg]^{2}\Bigg[\mid b_{i}\mid^{2}\int_{0}^{\pi}\pi^{2}_{i}\sin\theta d\theta+\mid a_{i}\mid^{2}\int_{0}^{\pi}\tau^{2}_{i}\sin\theta d\theta\Bigg]
\end{equation}

Using the values of integration from Equation (\ref{A40}) and Equation (\ref{A44}) in Equation (\ref{A46}), we have

$$Q_{sca\perp}=\frac{1}{x^{2}}\sum_{n=1}^{\infty}\Bigg[\frac{2n+1}{n(n+1)}\Bigg]^{2}\Bigg[n(n+1)\mid a_{n}\mid^{2}+\frac{n(n+1)(2n^{2}-1)}{2n+1}\mid b_{n}\mid^{2}\Bigg]$$
\begin{equation}\label{A49}
=\frac{1}{x^{2}}\sum_{n=1}^{\infty}\Bigg[\frac{(2n+1)^{2}}{n(n+1)}\mid a_{n}\mid^{2}+\frac{(2n+1)(2n^{2}-1)}{n(n+1)}\mid b_{n}\mid^{2}\Bigg]
\end{equation}

Also the using values of integration from Equation (\ref{A40}) and Equation (\ref{A44}) in Equation (\ref{A48}), we have

$$Q_{sca\parallel}=\frac{1}{x^{2}}\sum_{i=1}^{\infty}\Bigg[\frac{2i+1}{i(i+1)}\Bigg]^{2}\Bigg[i(i+1)\mid b_{i}\mid^{2}+\frac{i(i+1)(2i^{2}-1)}{(2i+1)}\mid a_{i}\mid^{2}\Bigg]$$
\begin{equation}\label{A50}
=\frac{1}{x^{2}}\sum_{i=1}^{\infty}\Bigg[\frac{(2i+1)^{2}}{i(i+1)}\mid b_{i}\mid^{2}+\frac{(2i+1)(2i^{2}-1)}{i(i+1)}\mid a_{i}\mid^{2}\Bigg]
\end{equation}

The total extinction efficiency (Van de Hulst 1981) can be expressed as,

\begin{equation}\label{A51}
Q_{ext}=\frac{4}{x^2}Re\{S(0)\}
\end{equation}

where
$$S_{1}(0)=S_{2}(0)=S(0)=\frac{1}{2}\sum_{n=1}^{\infty}(2n+1)(a_{n}+b_{n})$$

For $\theta=0^{\circ}$, the perpendicular component $S_{1}(0)$ and parallel component $S_{2}(0)$ of amplitude function are equal, so the perpendicular component $Q_{ext\perp}$ and parallel component $Q_{ext\parallel}$ of total extinction efficiency $Q_{ext}$ are also equal.
 
So from Equation (\ref{A51}), the expressions for $Q_{ext\perp}$ and $Q_{ext\parallel}$ can be expressed as follows\ :

\begin{equation}\label{A52}
Q_{ext\perp}=Q_{ext\parallel}=\frac{4}{x^2}Re\{S(0)\}=\frac{1}{2}\sum_{n=1}^{\infty}(2n+1)(a_{n}+b_{n})
\end{equation}

The single particle albedo (Van de Hulst 1981, Hapke 1993) can be expressed as :
\begin{equation}\label{A53}
w=\frac{Q_{sca}}{Q_{ext}}
\end{equation}

From Equation (\ref{A53}), the perpendicular and parallel components of single particle albedo can be defined as,
\begin{equation}\label{A54}
w_{\perp}=\frac{Q_{sca\perp}}{Q_{ext\perp}}
\end{equation}
\begin{equation}\label{A55}
w_{\parallel}=\frac{Q_{sca\parallel}}{Q_{ext\parallel}}
\end{equation}

By substituting Equations (\ref{A49}), (\ref{A50}) and (\ref{A52}) in Equations (\ref{A54}) and (\ref{A55}), $w_{\perp}$ and $w_{\parallel}$ can be now computed.

\section*{References}
\nocite{*}
\bibliography{mybibfile}

\end{document}